\documentclass[preprint,pra,aps,epsfig,showpacs]{revtex4}

\usepackage{graphicx}

\newcommand{\beq}{\begin{eqnarray}}
\newcommand{\eeq}{\end{eqnarray}}

\begin{document}
%\date{\today}
\setcounter{page}{1}

\title{Zero-temperature phase diagram of hard sphere bosons in asymmetric three dimensional optical lattices.}
\author{ F.~De~Soto, C. Carbonell-Coronado and M.C. Gordillo}
\affiliation{Departamento de Sistemas F\'{\i}sicos Qu\'{\i}micos y Naturales,
Universidad Pablo de Olavide. 41013 Sevilla, Spain}

\begin{abstract}
We studied the superfluid-to-Mott insulator transition for bosonic hard spheres loaded in asymmetric three-dimensional optical lattices
by means of diffusion Monte Carlo calculations.
The onset of the transition was monitored through the change in the chemical potential around  
the density corresponding to one particle per potential well.  
With this method, we were able to reproduce the results given in the literature for  
three-dimensional symmetric lattices and for systems whose asymmetry makes them equivalent to a set of quasi-one dimensional tubes.
%with optical potentials varying in the longitudinal direction. 
The location of the same transition for asymmetric systems akin to a stack of quasi-two dimensional lattices will be also given.   
Our results were checked against those given by a Bose-Hubbard model for similar arrangements. 
\end{abstract}


\maketitle

\section{Introduction}

An optical lattice is the result of the interference of a pair of laser beams to produce an standing wave. 
The change in the light intensity in different points of space create an effective periodic potential in whose minima  
neutral atoms can be confined \cite{gre2,bloch1,bloch2,greiner,bloch3}. By using several pairs of beams, the location 
of those small traps can be fixed to build periodic lattices of almost any geometry or dimension.
The most general form for a three dimensional optical lattice potential for an atom located 
at position $\vec{r}_i=(x_i,y_i,z_i)$, is:
\begin{equation} \label{pot}
V_{ext}(\vec{r}_i) = V_x \sin^2(k_x x_i) +  
 V_y \sin^2(k_y y_i) + V_z \sin^2(k_z z_i) 
\end{equation}
where each $k_n=2\pi/\lambda_n$ $(n=x,y,z)$ is related to a laser wavelength, $\lambda_n$,   
that in principle could be  different for any of the three space dimensions. $V_x,V_y$ and $V_z$ are the potential depths of the lattice minima that
can also differ from each other. Those depths are commonly given in units of the recoil
energy, $E_R= h^2/2 m \lambda^2$, ($h$ is the Planck constant and $m$ the mass of the neutral atom), and can be controlled experimentally by varying the 
intensity of the laser light. All this means that we can describe a dilute set of bosonic atoms (such as $^{87}$Rb \cite{gre2,paredes,
stoferle,spielman,clement,tr}, $^{133}$Cs \cite{gem,haller,mark} and $^{23}$Na \cite{imai})
of mass $m$ loaded in an optical lattice by the continuous Hamiltonian: 
\begin{equation}\label{hamiltonian}
H = \sum_{i=1}^N \left[ -\frac{\hbar^2}{2m} \triangle  + V_{ext} (\vec{r}_i) \right]  + \sum_{i<j} V(|\vec{r}_{ij}|)\ 
\end{equation}
in which $ V(|\vec{r}_{ij}|)$ stands for the interatomic potential between a pair of atoms $i$ and $j$ separated by a distance $r_{ij}$. 
In principle, one would have to solve the Sch\"odinger equation corresponding to the above Hamiltonian, but the most common approach has
been to simplify it to obtain the well-known Bose Hubbard \cite{jaksch,cazalilla} (BH) model. This is a discrete 
approximation that depends on two parameters
related to the one-body and two-body parts of the complete continuous Hamiltonian. 
The BH approximation assumes that
the interaction energy between two particles in the same 
optical well is negligible with respect to the energy difference between the two lowest Bloch states of the full periodic Hamiltonian. It also considers 
that the ground state of the system can be described by a superposition of functions localized entirely within each of those wells.     
Obviously, this implies that 
not all systems of neutral atoms on optical lattices could be described 
accurately by a Bose Hubbard Hamiltonian. 
In particular, that description is expected to break down 
for shallow potential depths ($V_x,V_y$ or $V_z$), in which both approximations fail \cite{bloch2}.  
On top of that, in the BH model the real interparticle interactions of the full Hamiltonian are further approximated by short-range pseudopotentials
that depend only on the scattering length and the mass of the atom.         
  
In this work, we will solve the full Schr\"odinger equation derived from the continuous Hamiltonian of Eq. (\ref{hamiltonian}), obtaining
the ground state of a set of  $N$ bosons loaded in the corresponding optical lattice. This means our results will be exact in the limit $T \rightarrow$ 0.
% One expects this ground state to be a very good description
%of the experimental setups given the extremely low temperatures at which those experiments take place \cite{bloch2,greiner}. 
Our only approximation will be to substitute the real atom-atom interaction by a hard-sphere potential, 
which is a common approach for homogeneous atomic gases \cite{boro1,gir,gregori,blume,pro}. 
This implies $V(|\vec{r}_{ij}|)=+\infty$ for $r_{ij}<a$ and 
$V(|\vec{r}_{ij}|)= 0$ for $r_{ij}>a$, with $a$ the diameter of the hard sphere fixed by the atomic scattering length.
With this in mind, we have only to define the corresponding $\lambda_n$ wavelengths to fully characterize the system we intend to study,  
in our case, $\lambda_x$ = $\lambda_y$ = $\lambda_z$ = 50 $a$.  
There is only a limited number of works that have used a continuous approximation  
for neutral atoms in optical lattices \cite{pilati,feli1,feli2,feli3}.  
Those have been devoted either to pure three-dimensional (3D) systems \cite{pilati} or to one-dimensional or quasi one-dimensional 
ones (1D) \cite{feli1,feli2,feli3}, i.e., they considered $V_x=V_y=V_z$ (3D) or $V_x = V_y >> V_z$ (1D) in Eq. (\ref{pot}). However, other combinations have
being experimentally produced as 3D-to-1D crossovers \cite{stoferle} and stacks of quasi two-dimensional (2D) optical lattices \cite{spielman}. 
Our goal will be to study the onset of the superfluid-Mott insulator transition in different asymmetric lattices.  

When the number of particles is not the same at each lattice minimum, the particles can jump to 
less populated wells, creating a superfluid phase.  We have also a superfluid when  
the number of neutral atoms is an integer multiple of the number of the lattice minima, but  
the depth of the optical lattice potential is low enough to permit tunneling between lattice sites. However, there is a critical value of those
depths for which tunneling becomes impossible. The phase in which the atoms are pinned to their respective positions 
is termed a Mott insulator (M.I.). In this work we will be dealing with Mott insulators for which the number of particles at each minimum is equal 
to one ($n$ = 1), even though phases with two or more particles per well are possible. 
%From all said above, we can deduce that 
The appearance of the Mott insulator can be
monitored then in two ways. First, for $n$ = 1, one can track the disappearance of the superfluid when the optical lattice potential increases. 
This is the way chosen in Ref. \onlinecite{pilati} for 3D systems. Second, we can check if the chemical potential of the system has a gap
around $n$ = 1. This is the approach used in simulations with the BH model \cite{batrou1,batrou2,laz} and in Refs. \onlinecite{feli1,feli2,feli3}. 
Ref. \onlinecite{feli3} tested that the superfluid and chemical potential criteria were equivalent for quasi one-dimensional systems of hard
spheres.  
            
The asymmetrical lattices were constructed by making $V_x= V_y = R V_z$, $R$ being a parameter that we varied in the range 0.01-100. 
In that way, we can model the results of some experiments in which the magnitude of the 
potential well in one direction is different than in the other two 
(see for instance Refs. \onlinecite{paredes,stoferle,clement,haller}).  In the case $R$= 1, 
we have a standard three dimensional symmetric lattice.  However, when $R \rightarrow$ 0, the 
atoms are trapped within quasi-two dimensional layers parallel to the $xy$ plane with no possibility of hopping in the $z$ direction  unless 
$V_z$ is small enough. In the opposite limit, $R \rightarrow \infty$, $V_x$ and $V_y$ are both very large with respect to $V_z$, 
allowing the movement of the confined species only in the $z$ direction.  
%In all those systems, the appearance of a Mott insulator phase will be monitored in a similar way than in previous works 
%\cite{feli1,feli2,feli3,batrou1,batrou2,laz}: 
For each $R$, there is a minimum value of the $V_z$ parameter (for $R >> 1$) or $V_x$, (for $R << 1$) from which we start to observe a discontinuity in the chemical potential  
when there is exactly one atom at each potential well. That critical value marks the onset of the superfluid-Mott insulator transition.    
For each simulation,  we fixed the value of $V_z$ and changed $R$ to produce different values of $V_x$ and $V_y$. However, we checked that 
proceeding the other way around, i.e., keeping constant $V_x$ and $V_y$ and making $V_z = V_x/R$, gave us the same values of the critical parameters.     

\section{Method}

To solve the Sch\"odinger equation we used a standard diffusion Monte Carlo (DMC) technique \cite{boro94}. This stochastic method provides 
an exact solution for the ground state of any system of bosons, within the uncertainties derived from the introduction of an initial approximation
to that ground state, the so-called {\em trial} function. The one chosen in this work was:    
\begin{equation}
\Psi_T({\bf r}_1,{\bf r}_2,\cdots,{\bf r}_N) = \prod_{i=1}^{N} \psi({\bf r}_i) \prod_{i<j=1}^{N} \phi(r_{ij})
\end{equation}
where the ${\bf r}_i$'s stand for the positions of the $N$ neutral atoms loaded in the optical lattice, and the $r_{ij}$'s correspond to the distance
between the pair of particles labeled $i$ and $j$. 
Here, $\psi({\bf r}_i)$ is the exact wavefunction for one particle in the external optical lattice potential,    
obtained by solving numerically the corresponding one-particle eigenvalue problem. The two-body correlations were taken into account through 
\cite{boro1}
\begin{equation}\label{trial}                                    
\phi(r_{ij})=\left\{
\begin{array}{lr}
0 &  r_{ij} < a\\
B \frac{\sin(\sqrt{\epsilon} (x-a))}{x} & a<r_{ij}<D,  \\ 
1-A e^{-x/\gamma} & r_{ij}>D 
\end{array} 
\right. 
\end{equation}
In principle, this expression depends on five constants, ($A$, $B$, $\epsilon$, $\gamma$ and $D$). However, by imposing the continuity of $\phi(r_{ij})$ and its first and second derivatives at $r_{ij}=D$ we are left with 
only two parameters to be obtained variationally.  
The primary output of a DMC calculation is the energy of the ground state. 
As we are working at $T=0 K$, that means that the chemical potential can be obtained straightforwardly from DMC results as the the derivative of the total energy of the system with respect to the number of particles. 
%The existence of a gap in the chemical potential at integer filling will be the sign of the existence of a Mott insulating (M.I.) phase.

\section{Results}

\begin{figure}
\begin{center}
\begin{tabular}{c}
\includegraphics[width=0.44\textwidth]{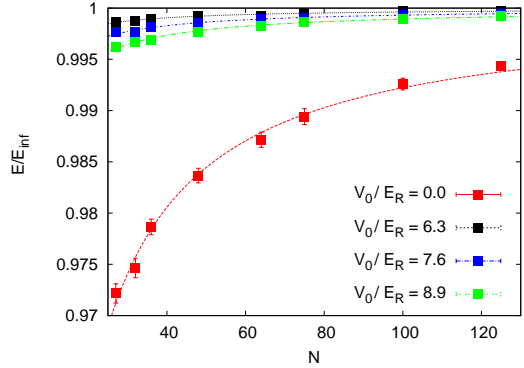} %\\
\end{tabular}
\caption{((Color online) Energy per particle for a system of $N$ particles divided by the same energy in the thermodynamic limit ($N \rightarrow \infty$)
as a function of $N$ for different values of the optical potential depth, $V_z$, for the case $R$= 1. See further explanation in the text. 
}
\label{fig:E_vs_N}
\end{center}
\end{figure}

In order to evaluate the stability of the results obtained
when we vary the size of the simulation cell used in our Monte Carlo calculations,
we have computed the energy per particle of a system of $N$ bosons in a symmetric three dimensional simulation cell with $N$ optical lattice wells, with different values of $N$ and different geometries of the simulation cell. We have done this for four values of the potential depth, $V=0$ and other three around the onset of the M.I. phase (see below). We considered simulation cells ranging from $3^3$ (i.e. 3 cells in each spatial direction) to $5^3$ and a corresponding number of particles ranging from 27 to 125. In Fig.~\ref{fig:E_vs_N} we displayed the ratio between the energy per particle for systems with a finite number of particles, $N$ ($E_N$), and that energy in the limit $N \rightarrow \infty$ ($E_\infty$), extracted for each potential depth from a fit 
of the type $E_N = E_\infty + b/N$. For $V_z$ = 0 (an homogeneous dilute gas of hard spheres), the $E_\infty$ derived from the
fit is fully compatible with the one obtained from Bogoliubov's perturbative analysis \cite{PhysRev.106.1135,PhysRev.115.1390} for the
density $na^3=6.4 \cdot 10^{-5}$, the numerical density corresponding in our simulation cells to one particle per potential well.

What we see in Fig. \ref{fig:E_vs_N} is that the finite size effects that could affect the energy are greatly reduced when one goes from 
an homogeneous gas to a gas loaded in an optical lattice. For instance, to use $E_{27}$ as approximation for $E_\infty$ for an homogeneous gas has an 
error of about 3\%, 
while to do the same for any of the $V_z$'s considered, reduces that same error an order of magnitude, up to $\sim$ 0.3 \%. If we use $E_{64}$ instead, 
the approximation is true within a 0.1\%. That is the reason why in most our calculations we considered simulation cells with a number of potential 
wells in the range $N$ = 27-64. Moreover, we verified that for $R$ = 1, the critical value of $V_z$ obtained from calculations with 
$3^3$, $4^3$ and $5^3$ simulation cells was the same. 
%3 $\times$ 3 $\times$ 3, 4 $\times$ 4 $\times$ 4 and  5 $\times$ 5 $\times$ 5 supercells was the same. 

\begin{figure}
\begin{center}
\includegraphics[width=8.5cm]{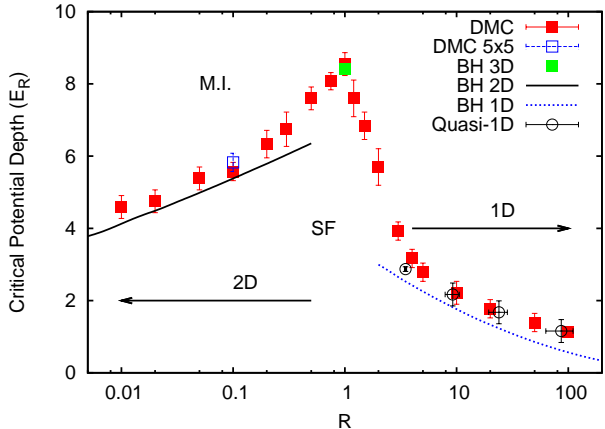}
\caption{(Color online) Critical value of the optical lattice depth responsible for the superfluid to Mott insulator phase transition (the minimum between $V_x$, $V_y$ and $V_z$) in units of the recoil energy 
versus $R=V_x/V_z$. Bose-Hubbard predictions in the limiting cases and quasi-one-dimensional results from Ref. \onlinecite{feli3} have been included for comparison.}
\label{fig2}
\end{center}
\end{figure}

The critical potential depths for the appearance of a Mott insulator phase obtained from our DMC calculations are displayed in Fig. \ref{fig2}, 
as a function of $R$, as full squares, together with their associated error bars.  
%The critical value for the appearance of the M.I. phase 
%is the smallest value of the potential depths for which the slope of the energy vs the particle density
%changes at one atom per potential well  \cite{feli1,feli2,feli3}. 
%Our DMC results
%are displayed as full squares in Fig.\ref{fig2}, together with their corresponding error bars. 
They were obtained for simulation cells in the range
$3^3$ ($R > 1$) to $4^3$ ($R < 1$) potential minima.
In that figure, for large values of $R$ (the right-hand end), our system 
can be described as a set  quasi-one dimensional tubes, since for $R \rightarrow \infty$, we can write Eq.~(\ref{pot}) as:  
\begin{eqnarray} \label{pot1d}
V_{ext}(\vec{r}_i) = R V_z k^2 (x_i^2 + y_i^2) + V_z \sin^2(k z_i) \nonumber \\
=  \frac{1}{2} m \omega_{\perp}^2 (x_i^2+y_i^2)  + V_z \sin^2(k z_i),
\end{eqnarray}
simply by developing $\sin(k x_i)$ and $\sin(k y_i)$ around $x = 0$ and $y = 0$. Thus, the critical behavior of these
optical lattices is regulated by $V_z$, the smallest of the triad $V_x,V_y$  and $V_z$. 
With that in mind,   
we calculated the critical $V_z$ parameters for systems with $R > 1$, and displayed them in Fig. \ref{fig2}. 
Since the optical lattice potential in the perpendicular direction can be approximated in this limit by a Gaussian, 
we can use the standard deviation of that function, $\sigma_{\perp}= (\hbar/m \omega_{\perp})^{1/2}$, as a measure of the tube "width". 
For our quasi one-dimensional systems, we have:
\begin{equation} \label{sigma1}
\sigma_{\perp} = \frac{\lambda}{ 2 \pi} \left( \frac {E_R} {R V_z} \right )^{1/4} 
\end{equation}
This means that, for $R= 100$ and $V_z/E_R \sim$ 1 (see, Fig. \ref{fig2}), we have $\sigma_{\perp} \sim 2.5 a$, a fairly thin tube, considering
that the distance between those quasi one-dimensional arrays is $\lambda/2$ = 25$a$.  
In Fig.~(\ref{fig2}) we show also (as open circles) the results of 
Ref. \onlinecite{feli3} that can be compared to the ones in this work. Those correspond to quasi-one dimensional Hamiltonians whose
optical lattice potentials are defined exactly by Eq.~(\ref{pot1d}), and were obtained using simulation cells with 40 potential minima, which are big enough to be  
free of finite size effects. The error bars in the $x$ axis for this set of points are derived from the equivalence $\frac{1}{2} m \omega_{\perp}^2 = R V_z k^2$, being 
$V_z$ the critical value for the transition in a quasi-one dimensional optical lattice. 
What we observe is that for $R \ge$ 10, both series of data follow exactly same trend, what implies that even a $3^3$ simulation
cell is big enough to produce reliable information for this kind of systems. The divergence found for $R\lesssim 3$ probably indicates the breakdown
of the approximation given in Eq. \ref{pot1d}. We also observe that for $1 < R < 3$ the $V_z$ critical values vary smoothly up to the  
one for a completely symmetric ($R$ = 1) 3D lattice ($V_z/E_R$ =  8.6 $\pm$ 0.3). This means that critical values for $R$'s other than the ones presented here
can be estimated reliably by interpolation between known values. In any case, the comparison of our data with the experiments is straightforward since our
results depend on the same magnitudes ($V_z$, $E_R$, $\lambda$ and $a$).   

If we go now to the left-hand side end of Fig. \ref{fig2} ($R \rightarrow$ 0), we end up with the following approximation to Eq.~(\ref{pot}): 
\begin{eqnarray} \label{pot2d}
V_{ext}(\vec{r}_i) =  V_x \sin^2(k_x x_i) + V_x \sin^2(k_y y_i) + \frac{V_x}{R} \sin^2(k z_i) \nonumber \\
=  V_x \sin^2(k_x x_i) + V_x \sin^2(k_y y_i)  +  \frac{1}{2} m \omega_{z}^2 (z_i^2),
\end{eqnarray}
obtained in the same way as before 
and taking in mind that now $\frac{V_x}{R}$ is very big.   
Eq. (\ref{pot2d}) corresponds to a set of quasi two dimensional systems  
whose critical behavior is regulated by $V_x=V_y$, the smallest parameter of the triad $V_x,V_y,V_z$. This value of 
$V_x$ is the one represented in the left-hand side of Fig. \ref{fig2}, not the $V_z$ value of the right-hand side already 
considered above. 
The limit $R\to 0$ corresponds to a strictly 2D system. 
The width of the Gaussian in the $z$ direction (the one in which the optical potential can be approximated by an harmonic potential) is: 
\begin{equation} \label{sigma2}
\sigma_{z} = \frac{\lambda}{ 2 \pi} \left( \frac {R E_R} {V_x} \right )^{1/4}, 
\end{equation}
what implies that for $R$ = 0.01 and $V_x/E_R \sim$ 4.5,  we have $\sigma_{z} \sim$ 1.7$a$, i.e., a set of a very thin quasi two-dimensional "pancakes".

In our simulations,  we used $3^3$ and $4^3$ supercells, checking that the results were similar in both cases (full squares), except for $R=0.1$ (open square), for which we represent also the critical $V_x$ value for a 
$5^2$ quasi-two dimensional lattice that obeys strictly Eq.(\ref{pot2d}). Both results are equivalent, i.e., they are not dependent on the geometry of the simulation cell, as in the case $R> 1$ discussed above. 
We can see also a smooth decay in the critical parameter from $R$ = 1 to the smallest $R$ considered ($R=$ 0.01), a behavior similar to the already discussed for
the quasi-one dimensional part of the diagram.
Unfortunately, we were not able to find in the literature any studies of neutral bosons loaded on 
quasi two-dimensional systems to compare these results to.          

However, they can be compared to the ones given by Bose-Hubbard model. This is a discrete model in which the positions
of the particles are limited to the bottom of each potential well. 
The BH Hamiltonian depends on two parameters, $J$, the hopping matrix element between nearest-neighbor sites, and $U$, related to the
interaction between two or more atoms in the same potential well. Both parameters can be obtained from the Hamiltonian in Eq. (\ref{hamiltonian}) 
by approximating the wave function of the system by a combination of Wannier functions corresponding to the lowest energy 
band \cite{jaksch,cazalilla}. In the limit $V_0 >> E_R$, we have 
\begin{equation} \label{J}
\frac{J}{E_R} = \frac{4}{\sqrt{\pi}} \left(\frac{V_0}{E_R}\right)^{\frac{3}{4}} e^{-2 \sqrt{\frac{V_0}{E_R}}}
\end{equation}
where $V_0$ is the potential depth that determines the critical behavior ($V_x$ or $V_z$). When, in addition to that approximation, 
we assumed that the interparticle interaction is described by a pseudopotential, we reach the following expression for $U$:     
\begin{equation} \label{U}
\frac{U}{E_R} =  \sqrt{\frac{8}{\pi}} ka \left(\frac{V_x}{E_R}\right)^{\frac{1}{4}} 
\left(\frac{V_y}{E_R}\right)^{\frac{1}{4}} \left(\frac{V_z}{E_R}\right)^{\frac{1}{4}}
\end{equation}
Both Eq.(\ref{J}) and Eq.(\ref{U}) are approximations, albeit widely used. %$k$ and $a$ have the same meanings as before.
%In principle, those equations they can be applied for any $V_x$, $V_y$ and $V_z$ triad, translating it into a  
%couple of $U,J$ values. However, we can operate backwards,  dividing  

From the above equations and the critical values of $U/J$ obtained from exact quantum Monte Carlo calculations found in the literature we can solve for the critical $V_0$ value  
for three-dimensional systems (where $U/J$=29.34, according to \cite{3HD}), two-dimensional ones (with $U/J$=16.739 from\cite{2HD}) and one-dimensional optical lattices 
(where $U/J$ is in the range 3.289-4.651, \cite{feli3,kuh}).
For $R >>1 $, this corresponds to 
the dashed line in Fig. \ref{fig2}, drawn using $(U/J)_{1D}$=3.97, (the center of the interval given above). 
What we find is that, in general,  the BH model underestimates the $V_z$ critical, even though when 
$R > 20$ the parameters deduced are closer to our simulation results. In fact, they could be even compatible when we consider all the 
$(U/J)_{1D}$ values in the literature \cite{feli3}. The same can be said of the opposite limit ($R <<1$, full line in Fig. \ref{fig2}): 
the BH results are always below that of the continuous Hamiltonian, but within the error bars of our data for $R <$ 0.1.  
On the other hand, the agreement between our $R$ = 1 result and the one derived from a 3D BH model is very good ($V_z/E_R$ = 8.6 $\pm$ 0.3 versus
$(V_z/E_R)_{BH}$ = 8.4), both values being comparable to the experimental one of Ref. \onlinecite{mark} 
($V_z/E_R \sim$ 8.8, for a system with $\lambda/a \sim$ 
47). 
%All in all, we can say that the Bose-Hubbard description gives us smaller values of the critical parameters than 
%those for a system of hard spheres loaded in a similar asymmetrical optical lattice. 
%This is the same conclusion 
%as the one found in a similar study for symmetric 3D systems \cite{pilati}, but extended to quasi one-dimensional and quasi-two dimensional
%optical lattices. 
%In addition, 
Unfortunatelly, the BH model does not seem to describe the behavior in the range 0.2 $<R<$ 3, which is experimentally 
accessible for different values of $\lambda$. For instance, in  Ref. \onlinecite{gem}, $R$ is in the range [0.02-1.5]. 
Another example is the 3D-1D crossover regime of Ref. \onlinecite{stoferle}, where the parameter R goes from 1.42 to 5. 

\section{Conclusions}

We have performed DMC calculations that allowed us to calculate the critical potential depths for the appearance of a 
Mott insulator when the underlying optical lattices were asymmetrical ($V_x = V_y \ne V_z$), a quite common experimental setup. 
Those results can be compared straightforwardly 
to those of the experiments, since they depend on the same magnitudes ($V_x,V_y,V_z, \lambda$ and $E_R$), and not through indirect variables
as the $U$ and $J$'s in the Bose-Hubbard model.  In any case, we found that  
for quasi-one and quasi-two dimensional systems, 
our results were comparable but larger than the ones deduced from a Bose-Hubbard Hamiltonian. 
This is the same conclusion 
as the one found in a similar study for symmetric 3D systems \cite{pilati}, but extended to quasi one-dimensional and quasi-two dimensional
optical lattices. 

To be sure, we tested whether the $V_0/E_R$ critical values deduced from Eqs. (\ref{J}) and (\ref{U}), strictly valid when $V_0 >> E_R$, were different
when the $J$ and $U$ parameters were calculated numerically from the exact solution of the one-body problem taken  
from Ref. \onlinecite{jaksch}.  
From those $J$ and $U$, the critical $V_z$ parameter for $R= 100$ was found to be 0.60$E_R$, versus 0.55$E_R$ deduced from Eqs. (\ref{J}) and
(\ref{U}) and the value displayed in Fig. \ref{fig2}. This is still lower than our continuous Hamiltonian result ($V_z/E_R \sim$ 1). On the other hand, 
for $R = 0.01$, the critical $V_x = V_y$ estimate was 4.14$E_R$, virtually identical to the number displayed in Fig. \ref{fig2} 
(4.13$E_R$). 
 
All our simulations deal with a single $\lambda/a$ parameter, but a comparison to previous calculations 
for totally symmetrical systems (Ref. \onlinecite{pilati}) or quasi-one dimensional ones (Ref. \onlinecite{feli3}), indicates that 
the critical $V_0/E_R$ values  obtained from realistic continuous Hamiltonians 
are still larger than the ones predicted by the BH model. There is also a rather smooth dependency of the critical value of $V_0$ on $\lambda/a$,
so in our asymmetrical lattices we would expect a qualitatively similar behavior than the already displayed in Fig. \ref{fig2}. 
In any case, we think that our results could provide a
guide for future experiments in atoms loaded in optical lattices. 

\acknowledgments

We acknowledge partial financial support from the 
Junta de Andaluc\'{\i}a group PAI-205 and grant FQM-5985, DGI (Spain) grant No. FIS2010-18356.

%\bibliographystyle{unsrt}
%\bibliography{refs}

\end{document}